\documentclass[journal=aelccp,manuscript=article]{achemso}
\usepackage{color}
\usepackage[version=3]{mhchem} 
\usepackage{pdfpages}

\author{Vikram Pande}
\affiliation{Department of Mechanical Engineering, Carnegie Mellon University, Pittsburgh, Pennsylvania 15213, USA}
\author{Venkatasubramanian Viswanathan}
\affiliation{Department of Mechanical Engineering, Carnegie Mellon University, Pittsburgh, Pennsylvania 15213, USA}
\email{venkvis@cmu.edu}

\title{Descriptors for Electrolyte-Renormalized Oxidative Stability of Solvents in Lithium-ion Batteries}

\usepackage{longtable}
\begin{document}

\begin{tocentry}
\includegraphics[width=7.0cm]{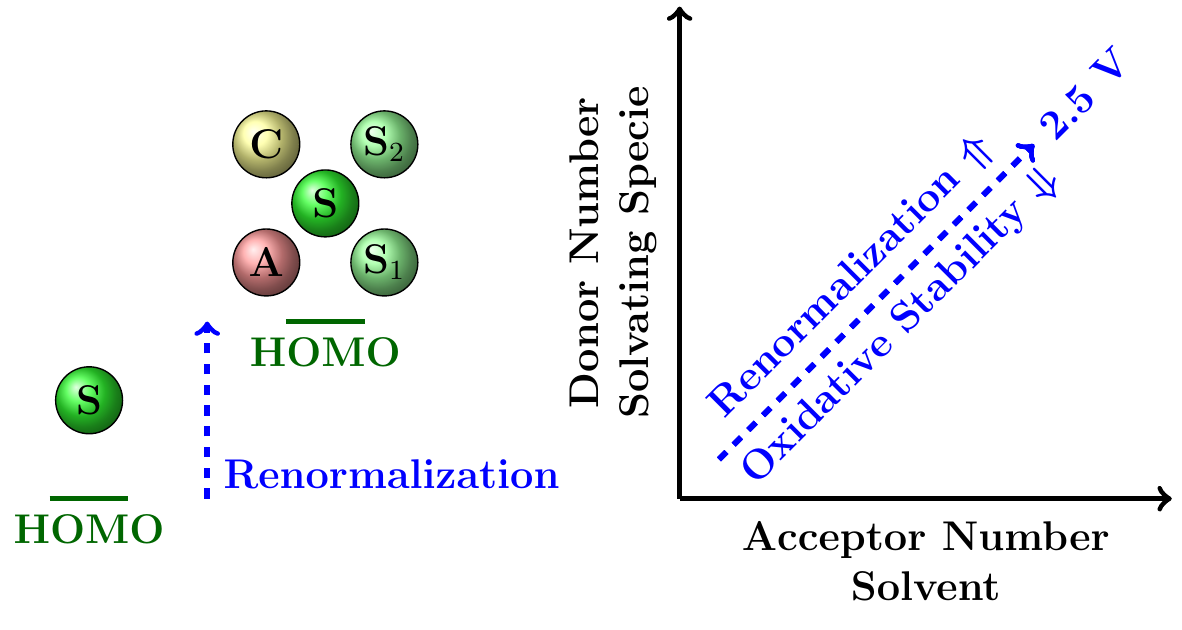}
\end{tocentry}

\begin{abstract}
 Electrolyte stability against oxidation is one of the important factors limiting the development of high energy density batteries. HOMO level of solvent molecules has been successfully used for understanding trends in their oxidative stability but assumes a non-interacting environment. However, solvent HOMO levels are renormalized due to molecules in their solvation shells. In this work, we first demonstrate an inexpensive and accurate method to determine the HOMO level of solvent followed by simple descriptors for renormalization of HOMO level due to different electrolyte components. The descriptors are based on Gutmann Donor and Acceptor numbers of solvent and other components. The method uses fast GGA-level DFT calculations compared to previously used expensive, experimental data dependent methods. This method can be used to screen for unexplored stable solvents among the large number of known organic compounds to design novel high voltage stable electrolytes. 
\end{abstract}

\newpage
The current state of the art lithium ion batteries do not meet the requirements for energy storage in terms of cost, energy content, cycle life and charging time. To meet the requirement for electric vehicle and grid storage applications, we need batteries which can deliver higher specific energy and energy density at cell level. To achieve such high specific energy and energy density, two major approaches have been employed to improve the current cathode: 1) Use of high voltage cathode materials such as layered LiNi$_x$Mn$_y$Co$_{z}$O$_2$ (NMC), spinel oxides such as LiNi$_{0.5}$Mn$_{1.5}$O$_4$ and poly-anion materials with Ni and Co cations such as LiCoPO$_4$, LiNiSO$_4$F, LiCoPO$_4$F, etc.\cite{etacheri2011challenges,kraytsberg2012higher} and 2) Use of high capacity conversion cathodes mainly S and O$_2$.\cite{abraham1996polymer,christensen2011critical,bruce2012li,evers2012new} One of the major factors limiting the use of high voltage cathode materials has been the lack of electrolytes which are stable against oxidation at these high cathode potentials.\cite{goodenough2009challenges,kim2014challenges,makeev2019computational} Li-O$_2$ batteries are also significantly affected by solvent degradation resulting from reactions with Li$_2$O$_2$ and O$_2^-$ existing in the cathode and electrolyte respectively.\cite{girishkumar2010lithium,mccloskey2012twin} The solvents oxidative stability is dependent on the environment including the all the electrolyte species and electrode surfaces.\cite{borodin2018challenges,khetan2017effect} Additives added to the electrolyte to form a good SEI at the anode, for non-flammability and increasing ionic conductivity influence the solvation shell of the solvent molecules.   Thus while designing new solvents and additives, we need to be ensure the additives do not affect the solvent's oxidative stability negatively. To search for new electrolyte components that are stable against oxidation, we need a fast and reliable method based on simple descriptors. 

\begin{figure}[!htb]
\centering
 \includegraphics[width=0.89\textwidth]{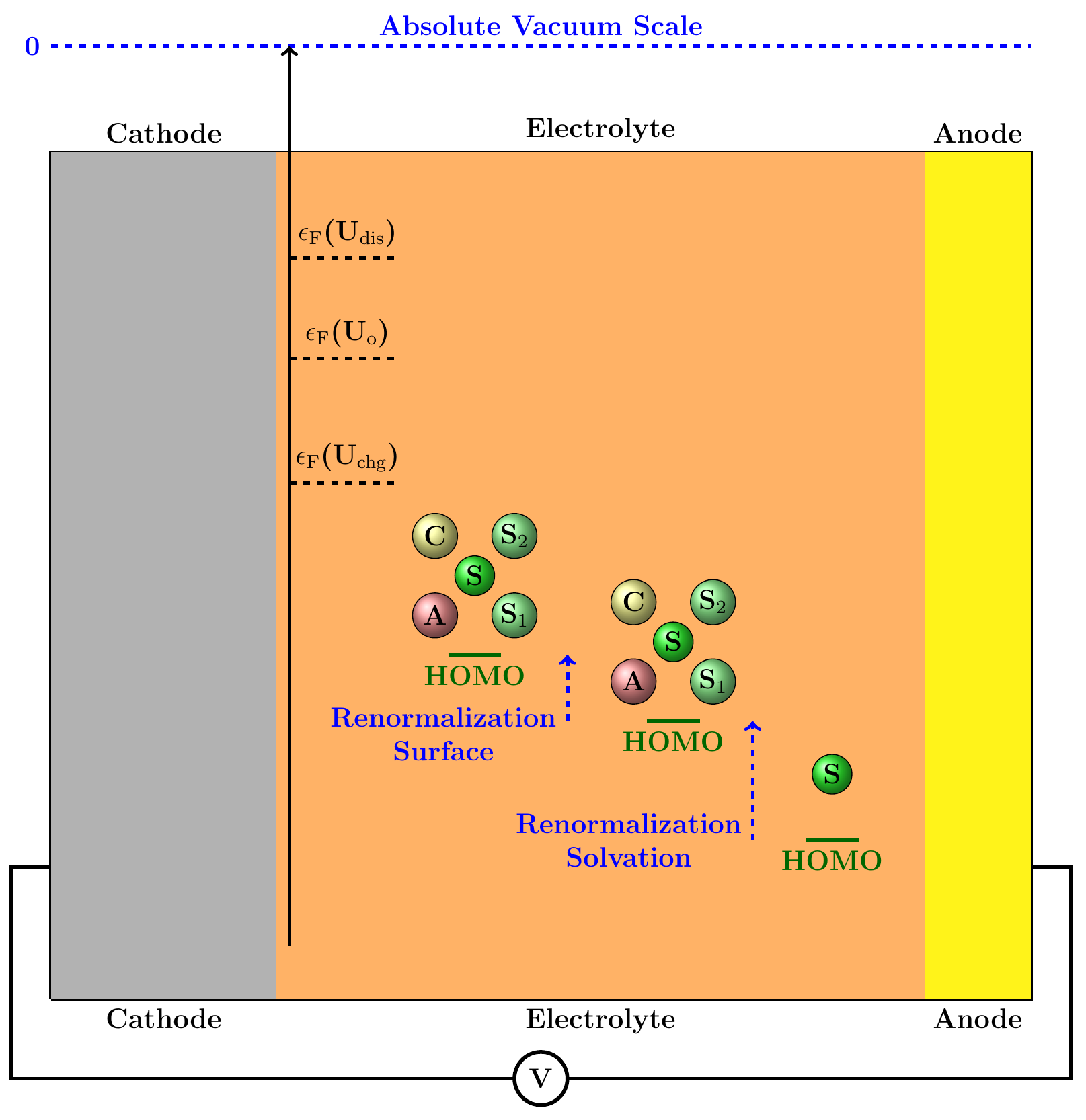}
 \caption{Schematic showing the renormalization of HOMO level of the solvent due to solvation and interactions with the cathode surface. The Fermi level of the cathode under charge-discharge is shown as reference for understanding stability.} 
 \label{schem}
\end{figure}

Oxidative stability of a molecule is primarily dependent on the relative position of its highest occupied molecular orbital level (HOMO) to the chemical potential of the cathode.\cite{goodenough2009challenges} HOMO level of the molecule specifically refers to the HOMO level in presence of molecule's neighbouring environment which includes the solvation species and the cathode surface.\cite{borodin2018challenges} Recent work has shown that for various battery electrolytes, oxidation involves interactions between multiple species such as the anions and other solvent molecules implying that the solvent's HOMO is renormalized  under solvation.\cite{peljo2018electrochemical,borodin2017modeling,borodin2015towards} Thus studying this renormalization due to different molecules becomes of utmost importance. Previous literature has also shown significant changes in the HOMO level of molecules at different cathode surfaces.\cite{kumar2014crystal,thygesen2009renormalization,giordano2017chemical} Previous studies have shown that renormalization of HOMO and LUMO levels of molecules near various cathode and metallic surfaces.\cite{neaton2006renormalization,thygesen2009renormalization,garcia2009polarization,khetan2017effect} The extent of renormalization at surfaces has been shown to be correlated with the surface and/or molecular density of states at the Fermi level.\cite{thygesen2009renormalization}

In this work, we focus on determining simple descriptors for the renormalization of HOMO levels due to solvation effects. Previous literature has used Quantum chemistry (QC) methods,\cite{borodin2013oxidative,alvarado2018carbonate} coupled cluster theory (CCSD(T)),\cite{jonsson2015electrochemical,kim2015computational} GW method\cite{khetan2017effect,thygesen2009renormalization} and DFT calculations employing polarizable continuum model for determining oxidative stability of electrolytes.\cite{kumar2014crystal,okoshi2015theoretical} These methods are either computationally expensive or require input parameters such as dielectric constant, cavity size, molar volumes and experimental Gibbs solvation energies.\cite{cossi2003energies,barone1998quantum,held2014simplified}  Here, we use simple GGA level DFT calculations and model solvation explicitly to determine the relation between the renormalization of the HOMO level of the solvent and the Gutmann Donor number (DN) \cite{gutmann1976solvent} and Acceptor number (AN)\cite{mayer1975acceptor} of the components in the electrolyte. Our method is computationally inexpensive and only relies on Donor and Acceptor number measurements compared to other methods. To identify new stable solvents, we first calculate the ionization potential of solvent with GGA-level DFT calculations. Using the model relation along with Donor and Accepetor numbers of solvents, salts and additives, we can easily estimate the renormalization of solvent's HOMO level in a given solvation environment. Our current method provides an easy way to incorporate electrolyte stability in high-throughput computational screening.\cite{makeev2019computational,bhowmik2019perspective}

 HOMO level is computed as the negative of the ionization potential (IP) as given by the Koopman's theorem.\cite{koopmans1934zuordnung} Self-Consistent DFT calculations were used to compute the IP of different solvent molecules with and without solvation. DFT calculations were done using the real space projector-augmented wave method as implemented in the GPAW code.\cite{mortensen2005real,enkovaara2010electronic}.  For all calculations, the converged force was $<$ 0.05 eV/\AA and a Fermi smearing of 0.01 eV was used.  Based on the calculations done by Rostgaard et. al., DFT energies of the neutral and charged solvent molecules can used to calculate IPs of solvents and solvated complexes with an accuracy better than the GW and MP2 methods which are very expensive computationally.\cite{rostgaard2010fully}  We use PBE exchange correlation functional\cite{perdew1996generalized} for the DFT calculations as it provides the highest accuracy in predicting IPs of the simulated list of solvents. 
The solvent molecules are first converged and relaxed in a vacuum space of 10 \AA \,in all directions. This is followed by the relaxation of the same molecule with a positive charge in a vacuum space of 10 \AA \,in all directions. The IP is calculated as the energy difference of the charged and neutral species. 
\begin{equation}
    IP = E_{DFT}(S^+) - E_{DFT}(S)
\end{equation}

\begin{figure}[!htb]
\centering
 \includegraphics[width=0.89\textwidth]{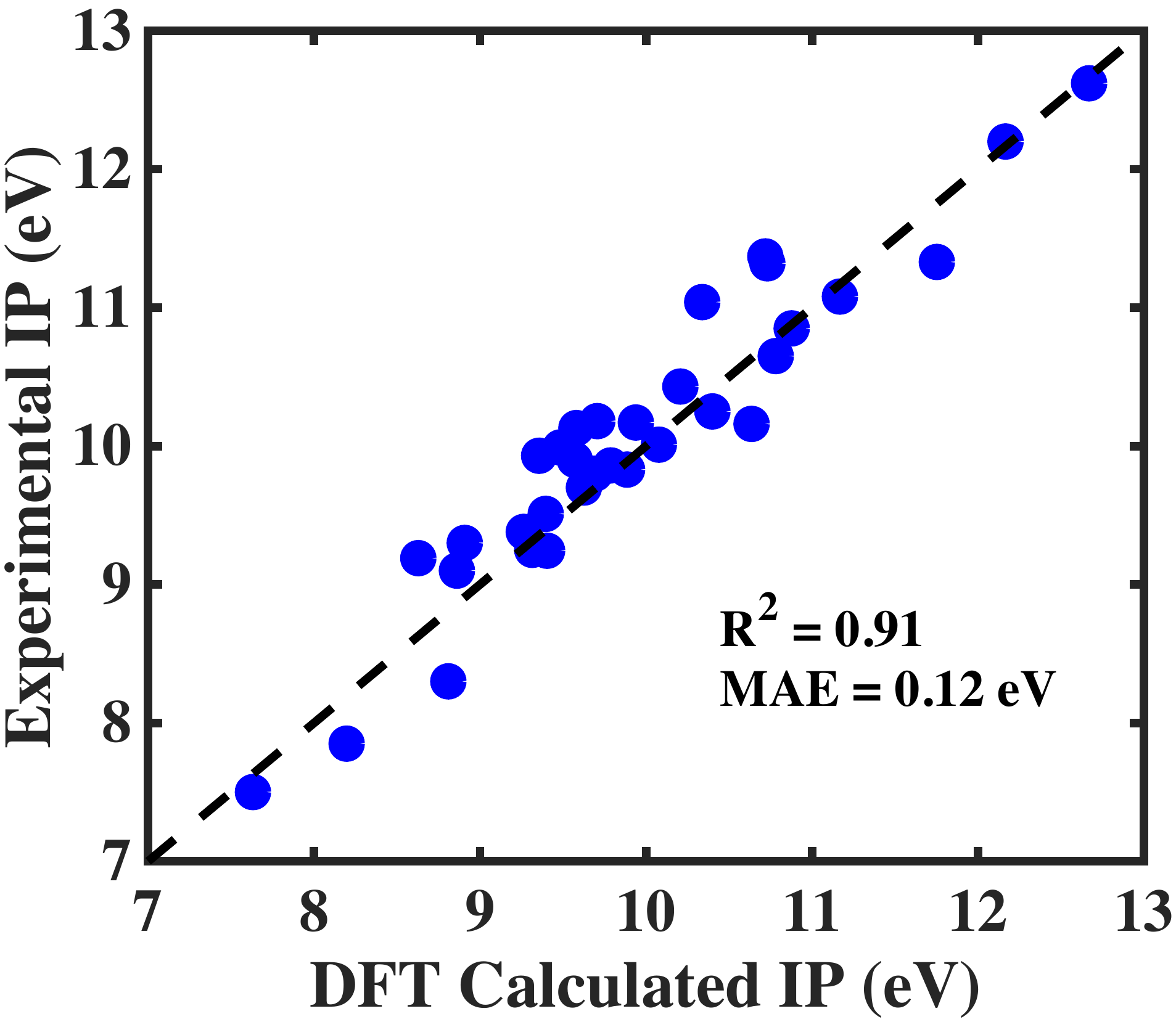}
 \caption{The correlation between the DFT calculated Ionization potential and experimental Ionization potential for 33 solvents.  We $R^2 = 0.91$ on the linear fit and an MAE of fit was 0.12 eV.}
 \label{ipneut}
\end{figure}

To validate the accuracy of our method of calculating IP, we calculated IPs for 34 solvents from the Stenutz dataset, with known experimental IPs. The DFT derived IPs agree well with the experimental values with a mean absolute error(MAE) of 0.12 eV as shown in Fig. \ref{ipneut}. Thus, this method should yield reliable estimates for IP of different molecules and complexes.

The next step is to calculate the IP of solvent in the solvation environment of cations and anions. As discussed before, it is known that the oxidative stability is reduced primarily due to anions and we will focus on this effect in this work.\cite{borodin2013oxidative} We have considered two cases, (1) The solvent is solvated by a single anion and (2) the solvent is solvated by the Li-anion ion pair. 


The complex of the given solvent-anion pair was constructed such that the geometry maximized the interaction between the negatively charged atoms of the anion and the acidic hydrogen atoms of the solvent molecule. This configuration was chosen because it has been shown previously that the anion-solvent interaction is primarily dominated by H-bond interactions.\cite{parker1962effects} For the solvent-ion pair complex, the configuration considered had the same solvent-anion structure with Li$^+$ added on the other side of the solvent. The complex structure was relaxed in a 10 \AA \,in all directions. A positive charge was placed on the complex and it was allowed to relax. As mentioned above, the energy difference of neutral and charged complex gave IP of complex. Bader charge analysis\cite{tang2009grid} was performed on the positively charged complex to confirm the oxidation of the solvent or the anion. To determine the partial charges, the vacuum charge density was set to zero in the Bader analysis. If the solvent is oxidized, then the renormalization is calculated as the difference in the IP of the the molecule and the solvated complex. Otherwise, the difference is calculated with respect to the IP of the anion.

To study the effect of solvation environment, we simulated a combination of 7 solvents: triethyl amine (TEA) (AN = 1.4 kcal/mol), 1-2 dimethoxy ethane (DME) (AN = 10.2 kcal/mol), dimethyl sulfoxide (DMSO) (AN = 19.3 kcal/mol), N-methylformamide (NMF) (AN = 32.1 kcal/mol), ethanol (AN = 37.9 kcal/mol), methanol (AN = 41.5 kcal/mol) and water (AN = 54.8 kcal/mol) and 7 anions: PF$_6^-$ (DN = 2.5 kcal/mol), ClO$_4^-$ (DN = 8.44 kcal/mol), TFSI$^-$ (DN = 11.2 kcal/mol), CF$_3$SO$_3^-$ (DN = 16.9 kcal/mol), NO$_3^-$ (DN = 21.1 kcal/mol), SCN$^-$ (DN = 31.9 kcal/mol) and OCN$^-$ (DN = 40.4 kcal/mol). The solvents were chosen to cover large range of ANs i.e interaction with anions and the salts had a large range of DNs of their anions. The Acceptor numbers for the solvents were taken from the  work by Mayer et. al.\cite{mayer1975acceptor} and the Donor number of anions were from the work by Linert et. al.\cite{linert1993donor}

\begin{table}
\caption{The Ionization potentials of the anions calculated using DFT with and without Li$^+$. The experimental Ionization potential (electron affinity of neutral specie) was taken from the NIST database. }

\begin{tabular}{|c|c|c|c|}
\hline
Anion	&	IP (eV)	&	IP with Li$^+$ (eV) & Experimental IP (eV)	\\
\hline
CF$_3$SO$_3^-$	&	5.0	&	10.0 & 5.3	\\
\hline
ClO$_4^-$	&	5.1	&	10.4 & 5.2	\\
\hline
NO$_3^-$	&	3.7	&	9.9 & 3.9	\\
\hline
PF$_6^-$	&	6.8	&	12.1 & -	\\
\hline
TFSI$^-$	&	5.2	&	9.5 & -	\\
\hline
OCN$^-$	&	3.7	&	9.3 & 3.6	\\
\hline
\end{tabular}
\end{table}

The IP of the anions was also calculated using the same method to determine their oxidative stability. By definition, the IP of the anion is negative of the electron affinity(EA) of the corresponding neutral specie. We find that the EAs compare well with experimental results as shown in Table 1. The IP of the anions is lower than all the considered solvents, which implies that the anions would likely oxidize much earlier than the solvent. However, the corresponding Li$^+$-anion ion pairs are more stable against oxidation compared to the solvents as seen in Table 1. Hence in general cations in the solvation shell will result in increase of oxidative stability of the molecule. The considered anions probably exist as ion-pairs and hence will be more stable against oxidation compared to the solvents.\cite{khetan2018understanding}

\begin{figure}[!htb]
\centering
 \includegraphics[width=0.89\textwidth]{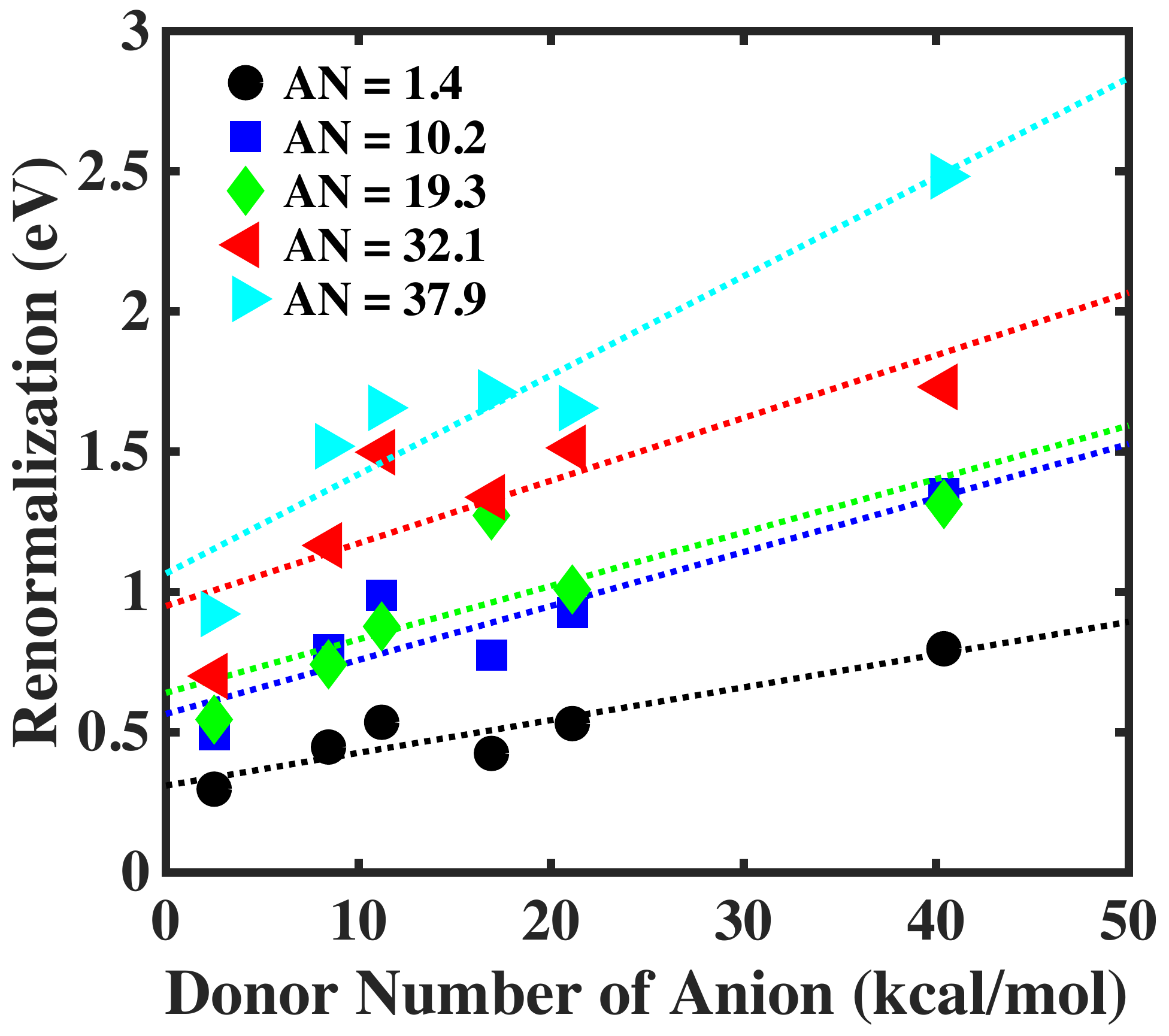}
 \caption{The renormalization of HOMO level of solvent for solvents with different Gutmann Acceptor numbers and salts with different Gutmann Donor numbers. The Acceptor numbers in the legend also have a unit of kcal/mol. The markers are the calculated data points from DFT and the dotted lines are linear fits.} 
 \label{renanion}
\end{figure}

For the Bader analysis, we consider the solvent completely oxidized if it has a net charge on it is greater than 0.6 and the anion completely oxidized if the charge on the anion is less than -0.4. All other cases are treated as co-oxidation of the anion and the solvent. Most anion ion-pairs oxidize before water and methanol due to their high ionization potentials, while in other cases solvents oxidize before the Li-anion ion-pairs. The calculated IP for the Li$^+$-SCN$^-$ ion pair is lower than most of the solvents considered and hence SCN$^-$ is oxidized before the solvent. For the cases where there is solvent oxidation, we expect that donacity of anions should help stabilize the oxidized solvent molecule and hence make oxidation more favorable. As shown in Fig. \ref{renanion}, we do see higher DN anions  reduce the IP of the solvent to a greater extent i.e. have greater renormalization of HOMO level of solvent. We also see that for a given salt, solvents with higher AN have larger reduction in IP. Thus, we can conclude that the DN of the anion and the AN of the solvent are strongly correlated to the oxidative stability of the solvent. 

Now to quantify the extent of this effect of renormalization due to solvation and avoid further DFT calculations, we explore two simple models based on these descriptors. For the first model, we assume that the renormalization is a simple linear function of the AN and DN given by:
\begin{equation}
    RN_{HOMO} = C + \alpha (DN) + \beta AN
\end{equation}
The second model is based on a geometric mean of the interaction strength. Here we assume that the renormalization is proportional to the electrostatic interaction energy between the two species. The DN is the electron donating tendency of a molecule and is related to the negative partial charge of the molecule. Similarly the AN is related to the positive partial charge of the molecule. Then the interaction energy could be correlated to the product of the DN and AN of the corresponding species. Among the considered models, geometric mean, given by:
\begin{equation}
    RN_{HOMO} = C + \alpha \sqrt{AN\times DN}
\end{equation}
gives the best fit for the data.

\begin{figure}[!htb]
\centering
 \includegraphics[width=0.89\textwidth]{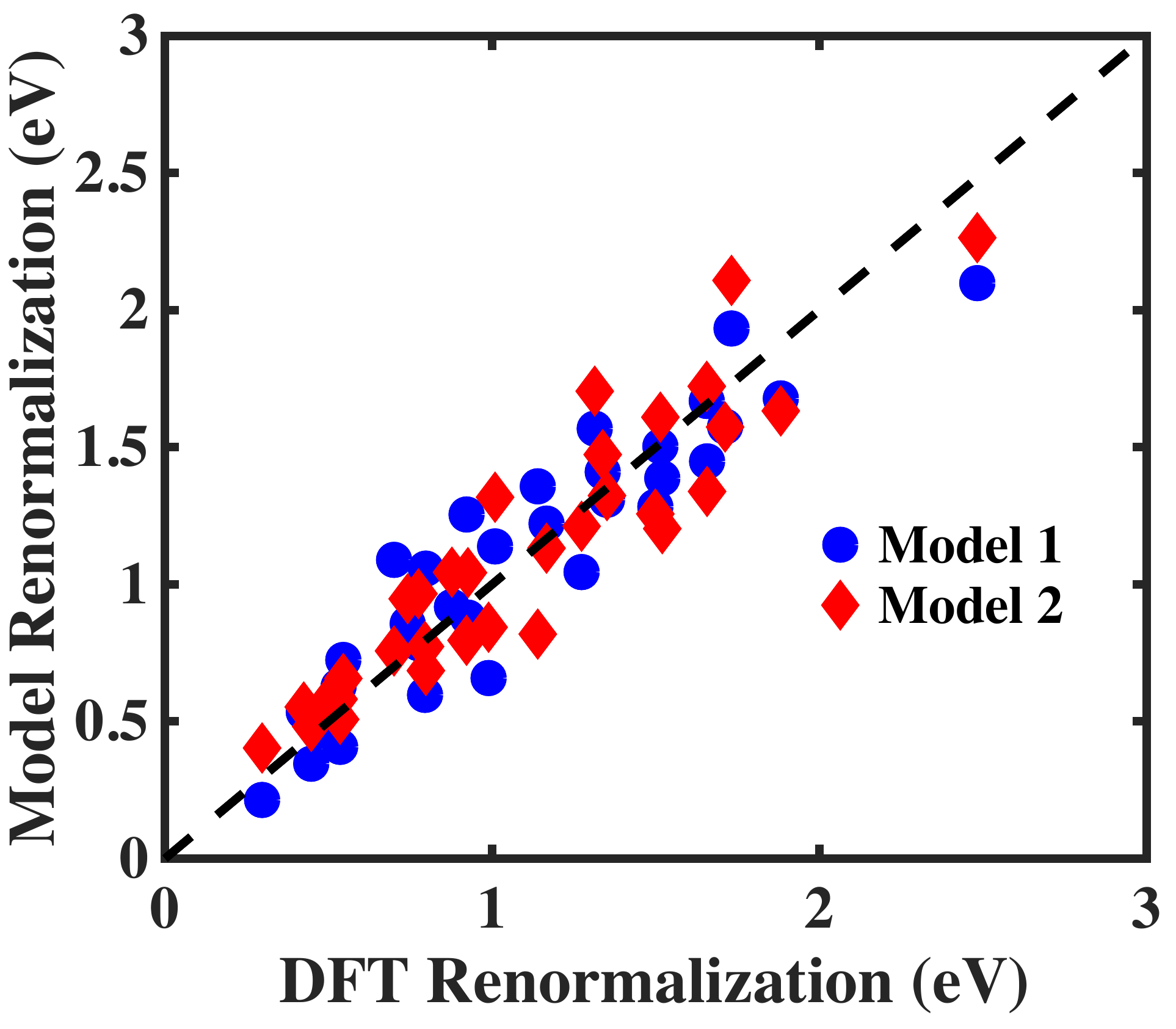}
 \caption{Comparison of DFT calculated values and descriptor based model predicted values. Model 1 assumes linear variation w.r.t to Acceptor and Donor numbers, while Model 2 assumes an electrostatic model resulting in the geometric mean of the Acceptor and Donor numbers as the composite descriptor. } 
 \label{moderr}
\end{figure}

 The coefficients for these models were trained using 32 DFT calculations covering binary combinations of 6 salts and 6 solvents mentioned earlier. Note that the salt and solvents cover a significant range of the known AN and DN space. Surprisingly, both models as shown in Fig. \ref{moderr}, perform equally well with an R$^2$ of 0.86 and a mean absolute error (MAE) of 0.16 eV. The goodness of the fit with experimental AN and DN also proves that for all the binary salt and solvent combinations, the number of species in the solvation shell of the salt anions is not significantly different. 
 
 An advantage of building such a simple descriptor-based model is that now we can also evaluate the renormalization of HOMO level of a solvent in presence of other solvent molecules if the DN and AN of the corresponding species are known. Determining solvent-solvent interactions through DFT is quite expensive and non-trivial due to the large number of conformers for most of these organic compounds. This model now provides a way to easily estimate this effect. Utilizing these results, we can now determine the net renormalization due to the entire solvation shell of the solvent molecule as shown in Fig. \ref{schem} by summing the individual specie contribution. Our previous work also shows a way to estimate the solvation shell composition for electrolyte mixtures.\cite{burke2015enhancing}

\begin{figure}[!htb]
\centering
 \includegraphics[width=0.89\textwidth]{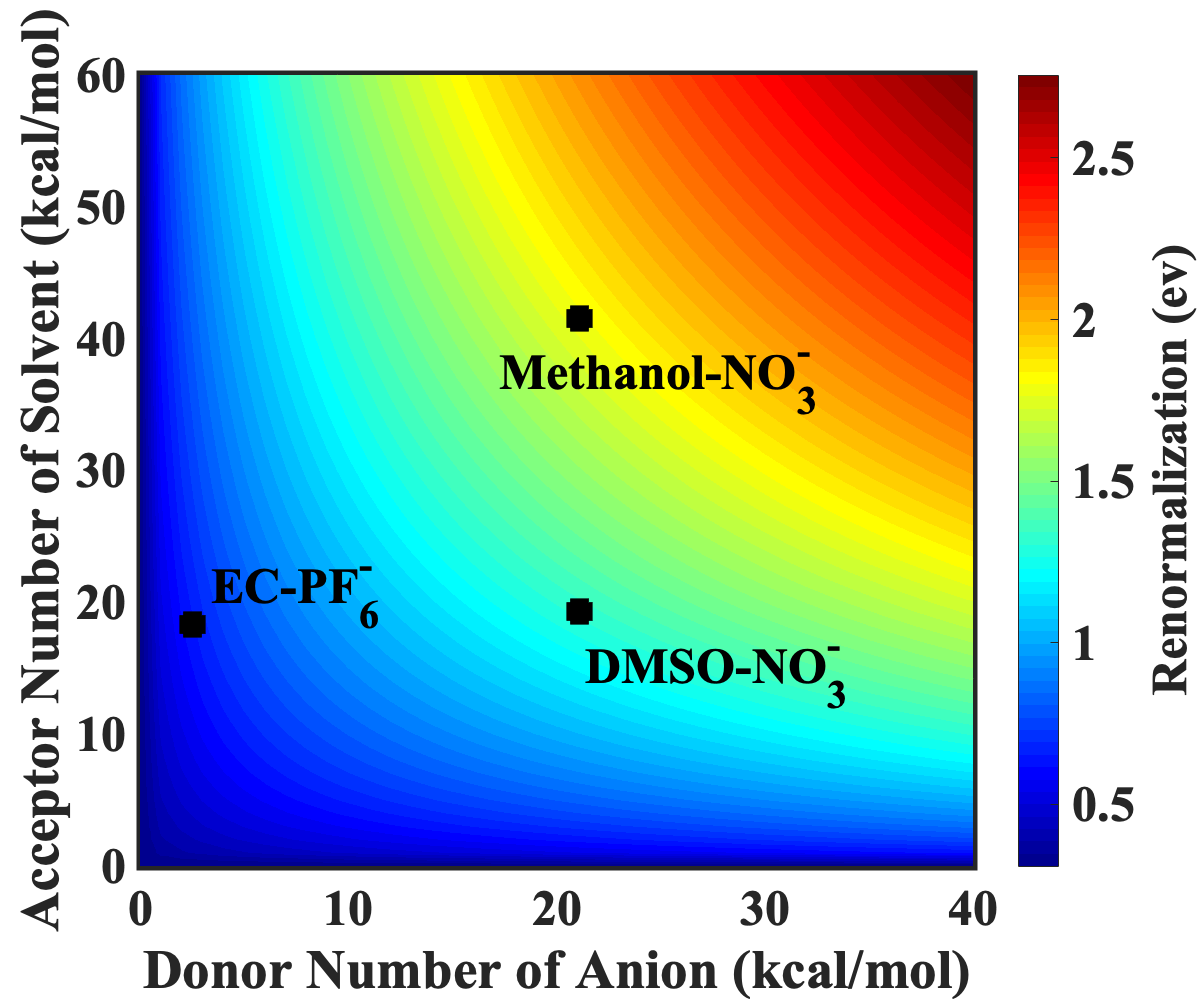}
 \caption{The renormalization of HOMO level of solvent for solvents as a function of Gutmann Acceptor number of solvent and Gutmann Donor number of salt as predicted by the models. The black dots represent some of the commonly used electrolytes in Li-ion and Li-O$_2$ batteries.} 
 \label{rencont}
\end{figure}

Using the models, we also derive a contour plot for easily calculating the renormalization as a function of the AN and DN of the respective species as shown in Fig. \ref{rencont}. In the standard Li-ion battery electrolyte comprising of EC and LiPF$_6$, the PF$_6^-$ would reduce EC's stability by 0.65 V and another EC molecule will reduce it by 1.14 V. These values are very close to the ones calculated using Quantum Chemistry based methods.\cite{borodin2018challenges,borodin2013oxidative}

We have shown in an earlier work that larger DN of the electrolyte salt anion and larger AN of the solvent leads to increased solvation and in turn increased solubility of the Li salt.\cite{burke2015enhancing, khetan2018understanding} However, we find that this leads to a reduction in oxidative stability of the solvent. So, to avoid the issue of reduction in stability, it is necessary to use low AN solvents such as ethers, esters, carbonates, etc and low DN salts such as LiPF$_6$, LiTFSI, etc.  This now provides a rational basis for modifying electrolyte constituents in current Li-ion batteries.

In the context of Li-O$_2$ batteries, the stability of the solvent will also be affected by the superoxide anion (O$_2^-$) which is generated during the discharge and charge processes. The DN of O$_2^-$ anion is expected to be greater than that of OCN$^-$ (40 kcal/mol). Now using the model, we estimate that a O$_2^-$ molecule would reduce oxidative stability of solvents by 1.25-1.75 V. Thus, for Li-O$_2$ we would need low AN solvents number which are stable atleast up to 4.5 V vs Li/Li$^+$. The other approach would be to reduce the amount of O$_2^-$ formed by addition of redox mediators which has already been demonstrated in literature.\cite{chen2013charging}

Lastly we would like to point out that all computation in this work are reference to vacuum. In battery literature, stabilities are measured on the Li/Li$^+$. As shown before, the Li/Li$^+$ redox potential also shifts by $\sim$ 0-0.75 V depending on the effective DN of the electrolyte mixture. This is another degree of freedom to expand the stability window of the electrolyte.


To summarize, we have identified simple descriptors to determine the influence of solvation on the oxidative stability of various electrolyte components. Our model can be used to quickly determine the oxidative stability of the electrolyte mixture once the DN (Lewis Basicity), AN (Lewis Acidity) and the IP of the individual components is determined. This method does not depend on other experimental measurements and hence can be used to evaluate unexplored solvent and salt molecules if we can develop a method to efficiently compute the DN and AN from first principles. We also believe utilizing a similar concept, it is possible to describe renormalization from the cathode surface as a function of the surface charge of the cathode surface\cite{garcia2009polarization}. We will explore this in a future study and complete the model for solvent stability against cathodes. Using our work we propose low AN and high DN solvents, low DN salts with low crystallization energies for getting high voltage stability while maintaining sufficient solubility of the Li salt.

\begin{acknowledgement}
 Acknowledgment is  made  to  the  Scott  Institute  for  Energy  Innovation at Carnegie Mellon for supporting V.P. during his graduate research. This work was also partially supported by the Assistant Secretary for Energy Efficiency and
Renewable Energy, Office of Vehicle Technologies of the US Department of Energy (DOE) through the
Advanced Battery Materials Research (BMR) Program under contract no. DE-EE0007810.

\end{acknowledgement}


\bibliography{refs.bib}

\includepdf[pages={1-}]{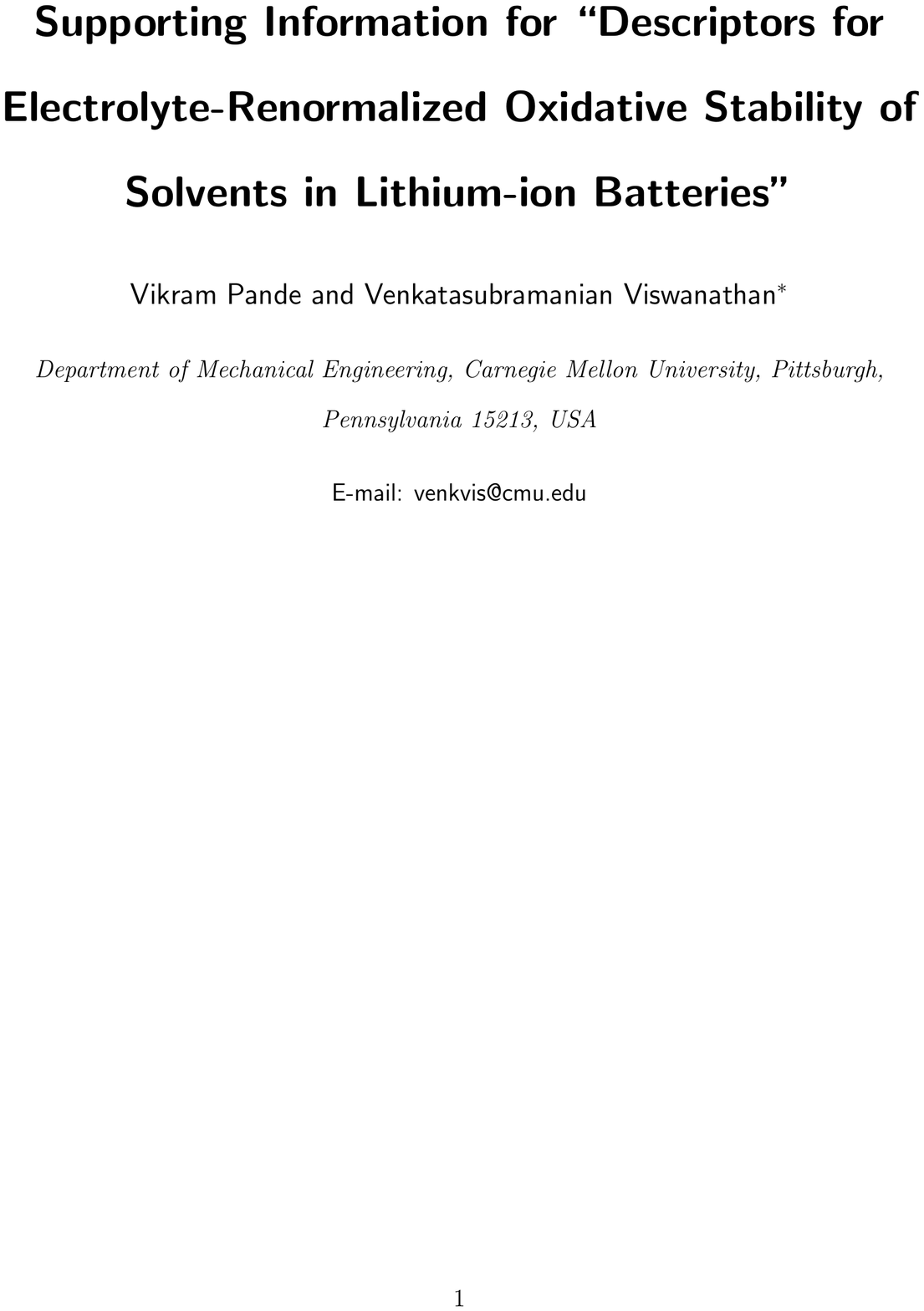}

\end{document}